\documentclass[aps,twocolumn]{revtex4}
\usepackage{amssymb}
\usepackage{amsmath}
\usepackage{mathrsfs}
\usepackage{graphicx}
\usepackage{dcolumn}
\usepackage{bm}

\begin{document}

\title{Gravity-induced transparency}
\author{Yongjie Pan}
\author{Baocheng Zhang}
\email{zhangbaocheng@cug.edu.cn}
\affiliation{School of Mathematics and Physics, China University of Geosciences, Wuhan
430074, China}

\begin{abstract}
We investigate the transition amplitudes of the Unruh-DeWitt detector within
the Schwarzschild spacetime background and discover gravity-induced
transparency phenomena akin to the earlier acceleration-induced
transparency. This is confirmed through calculations performed in both the
Hartle-Hawking and Unruh states. The similarity between free-falling
detectors in these states and accelerated detectors in electromagnetic
fields is noteworthy, as it parallels the scenario in which
acceleration-induced transparency phenomena occur.
\end{abstract}

\maketitle

\section{Introduction}

Acceleration-induced phenomena have garnered significant attention in
physics research. Notable examples include the Unruh effect \cite%
{unruh1976notes}, the anti-Unruh effect \cite%
{brenna2016anti,garay2016thermalization}, and acceleration-induced
transparency (AIT) phenomena \cite{vsoda2022acceleration}. While the former
two effects have been extensively studied theoretically, the AIT phenomenon
holds particular experimental significance. As is well-known, detecting the
Unruh effect experimentally presents challenges due to the exceedingly low
Unruh temperature; for instance, achieving a thermal bath at $1$ K would
require an acceleration of $10^{20}$ m/s$^{2}$. Recent work on AIT has
proposed a method to amplify the Unruh effect by subjecting the Unruh-DeWitt
(UDW) detector to acceleration within an electromagnetic field rather than
in vacuum. For the Fock state, the Unruh effect is enhanced by a factor of $%
n+1$ (where $n$ is the photon number) compared to the scenario occurring in
vacuum. Additionally, it has been found that the anti-Unruh effect can also
be amplified through a simulated light-matter interaction akin to the AIT
phenomenon \cite{pan2023enhanced}.

Given that the equivalence principle links the Unruh effect to the Hawking
effect \cite{hawking1974black}, and considering the challenge in detecting
Hawking radiation, it is pertinent to explore whether the AIT effect can be
extended to its gravity-induced analogue. Hawking radiation can be likened
to the electromagnetic field, while a detector near a Schwarzschild black
hole mirrors an accelerated detector in flat spacetime. Consequently, it
seems feasible to obtain gravity-induced transparency phenomena for a
detector near a black hole. However, a hurdle arises in obtaining analytic
expressions for field mode solutions \cite%
{hodgkinson2013particle,Hodgkinson2014Static} when calculating field mode
solutions in (3+1)-dimensional Schwarzschild spacetime. Despite this, the
transition rate of a detector near a black hole has been studied previously
in various scenarios, including static detectors near Schwarzschild black
holes \cite{Hodgkinson2014Static}, the anti-Hawking effect for BTZ black
holes \cite{henderson2020anti}, rotating BTZ black holes \cite%
{robbins2022anti}, and free-fall detectors near black holes \cite%
{scully2018quantum}. Hence, it is possible to apply the methodologies from
these studies to investigate gravity-induced transparency phenomena for a
UDW detector moving outside a Schwarzschild black hole.

This paper is organized as follows. In Sec. II, we present a model for the
interaction between detectors and fields in flat spacetime and extend the
definition of AIT phenomena. Sec. III is devoted to the investigation of
gravity-induced transparency phenomena for a free-falling detector in the
exterior region of a black hole. Finally, we summarize and give the
conclusion in Sec. IV.

\section{The interaction model}

We start with the model of the UDW detector, commonly regarded as a
pointlike two-level quantum system. The interaction Hamiltonian is expressed
as \cite{dewitt1979quantum,louko2008transition}
\begin{equation}
H_{I}=\lambda \chi (\tau )\mu (\tau )\phi \lbrack x(\tau ),t(\tau )],
\label{H}
\end{equation}%
where $\lambda $ is the coupling constant between the accelerated detector
and the scalar field, $\mu (\tau )=e^{i\Omega \tau }\sigma ^{+}+e^{-i\Omega
\tau }\sigma ^{-}$ represents the detector's monopole moment with $\sigma
^{\pm }$ being SU(2) ladder operators. The detector possesses two distinct
energy levels denoted by the ground $|g\rangle $ and excited $|e\rangle $
states, respectively, separated by an energy gap $\Omega $ in the detector's
rest frame. $\phi \lbrack x(\tau ),t(\tau )]$ is the field operator in which
$x(\tau ),t(\tau )$ represents the detector's trajectory, and $\chi (\tau )$
is the switching function. For the purposes of this paper, we set $\chi
(\tau )$ to be equal to 1.

The time evolution operator under the Hamiltonian (\ref{H}) is obtained
through the following perturbative expansion to the first order
\begin{align}
U& =1-i\int d\tau H_{I}(\tau )+\mathcal{O}(\lambda ^{2})  \notag \\
& =1-i\lambda \sum_{k}[\eta _{k}^{+}a_{k}^{\dag }\sigma ^{\dagger }+\eta
_{k}^{-}a_{k}\sigma ^{\dagger }+H.c.].  \label{eu}
\end{align}%
In the quantum description of light-matter interaction, $a_{k}^{\dag }\sigma
^{\dagger }$ and $a_{k}\sigma ^{-}$ are the counter-rotating wave terms,
while $a_{k}\sigma ^{\dagger }$ and $a_{k}^{\dag }\sigma ^{-}$ are the
rotating wave terms. $\sum_{k}$ represents the summation of all momentum
modes in the field. $\eta _{k}^{\pm }=\int \frac{\lambda d\tau }{\sqrt{(2\pi
)^{3}2\omega }}e^{i\Omega \tau \pm ik^{\mu }x_{\mu }}$ where $k^{\mu }x_{\mu
}=\omega t(\tau )-kx(\tau )$ are related to the trajectory of the detector. $%
a_{k}^{\dag }$ and $a_{k}$ are the creation and annihilation operators for
the field $\phi $ which can be expressed as $\hat{\phi}(x)=\int dk(u_{k}\hat{%
a}_{k}+H.c.)$ where $u_{k}=\left[ 2\omega (2\pi )\right] ^{-1/2}e^{-ik_{\mu
}x^{\mu }}$.

In the interaction picture, this evolution is described by
\begin{align}
U|g\rangle |n\rangle _{k}=& |g\rangle |n\rangle _{k}-i\sqrt{n}\eta
_{-}|e\rangle |n-1\rangle _{k}  \notag \\
& -i\sqrt{n+1}\eta _{+}|e\rangle |n+1\rangle _{k},  \label{e1} \\
U|e\rangle |n\rangle _{k}=& |e\rangle |n\rangle _{k}-i\sqrt{n+1}\eta
_{-}^{\ast }|g\rangle |n+1\rangle _{k}  \notag \\
& -i\sqrt{n}\eta _{+}^{\ast }|g\rangle |n-1\rangle _{k}.  \label{e2}
\end{align}%
where $|n\rangle _{k}$ is the Fock state of the field indicating the
presence of $n$ photons in the $k$ mode. In the equations (\ref{e1}) and (%
\ref{e2}), the second and third terms on the right side represent the
rotating-wave and counter-rotating-wave terms, respectively. For instance,
the second term in Eq. (\ref{e1}) signifies the absorption of energy by the
detector from the field, causing a transition from the ground state to the
excited state, commonly referred to as the stimulated absorption term \cite%
{vsoda2022acceleration,pan2023enhanced}. The third term in Eq. (\ref{e1})
accounts for the contribution of the Unruh effect, arising from the
accelerated motion of the detector within the photon field, known as the
stimulated Unruh effect term. Equation (\ref{e2}) can be interpreted
similarly.

At first, we consider the field to be in the vacuum where $n_{k}=0$ for any $%
k$ mode. When the detector is accelerated in the vacuum state with the
trajectory \cite{crispino2008unruh,ben2019unruh} 
\begin{equation}
x^{\mu }(\tau )=[\sinh (a\tau )/a,\cosh (a\tau )/a],  \label{ua}
\end{equation}%
we have the transition probablity, which is proportional to
\begin{equation}
|\eta ^{+}|^{2}=\lambda ^{2}\frac{2\pi /(\Omega a)}{e^{\Omega
/(k_{B}T_{U})}-1}.
\end{equation}%
This corresponds to a Bose-Einstein distribution at a temperature $T_{U}=%
\frac{a}{2\pi }$. This implies that the conventional Unruh effect \cite%
{unruh1976notes,unruh1984happens}, which arises from a uniformly
accelerating motion of the detector in vacuum, is equivalent to the detector
being immersed in a thermal bath at a temperature of $T_{U}$.

When we replace the vacuum field state with a single-mode Fock state $%
|n\rangle $, we obtain the transition probability
\begin{equation}
P\varpropto (n+1)|\eta _{+}|^{2}.  \label{pn1}
\end{equation}

It is evident that the transition probability of the detector is amplified
by a factor of $(n+1)$. This amplification could bring the transition
probability of the detector into an experimentally observable range, thereby
facilitating the experimental observation of the Unruh effect. Achieving
this has been challenging in previous studies \cite{bell1983electrons}.

As electromagnetically induced transparency \cite%
{fleischhauer2005electromagnetically} refers to the suppression of the
rotating-wave term in the model of light-matter interaction (the
counter-rotating wave term is neglected due to its violation of the
conservation of energy), the AIT phenomenon presents a similar mechanism. In
AIT, the rotating-wave term in the interaction between the detector and the
field is suppressed, while the counter-rotating wave term contributes to the
transition of the detector. This can be expressed mathematically as the
condition that $\frac{|\eta _{-}|}{|\eta _{+}|}\ll 1$.

\section{Gravity-induced transparency}

Now we turn to the situation of the curved spacetime and examine the
behaviors of the detector in the vicinity of a Schwarzschild black hole,
characterized by the metric
\begin{eqnarray}
ds^{2} &=&-\left( 1-\frac{2M}{r}\right) dt^{2}+\left( 1-\frac{2M}{r}\right)
^{-1}dr^{2}  \notag \\
&&+r^{2}(d\theta ^{2}+\sin ^{2}\theta d\phi ^{2}),  \label{metric}
\end{eqnarray}%
where $M>0$ is the mass of the black hole.

To investigate the interaction between a UDW detector and the field within
the framework of Schwarzschild spacetime (\ref{metric}), and to compute the
transition amplitude of the detector, the vacuum states are required. In
this paper, we will perform calculations using two distinct field states:
the Hartle-Hawking state \cite{Hartle1976Path} and the Unruh state \cite%
{unruh1976notes}.

We begin by considering the free-falling process of the detector, commencing
at a distance $R$ from the center of the black hole. The initial state of
the field is denoted by $|n\rangle _{S}$, where $S$ distinguishes between
the two different field states. Following the interaction between the
detector and the field, we derive the transition amplitude of the detector as%
\begin{equation}
\eta _{S}^{\pm }=\int d\tau e^{i\Omega \tau }\langle n\pm 1|_{S}\phi
_{S}(\tau )|n\rangle _{S}.  \label{etazf}
\end{equation}%
This \textquotedblleft $+$\textquotedblright\ term corresponds to the
counter-rotating wave term, signifying the emission of photons from the
field when the detector transitions from the ground state to the excited
state. Conversely, the \textquotedblleft $-$\textquotedblright\ term
corresponds to the rotating-wave term, representing the process in which the
detector absorbs the energy of a photon from the field while transitioning
from the ground state to the excited state.

In order to calculate the transition amplitude (\ref{etazf}), we have to
solve the field $\phi _{S}$. Mode solutions of the Klein-Gordon equation in
the Schwarzschild spacetime have the form:
\begin{equation}
u=\frac{1}{\sqrt{4\pi \omega }}r^{-1}\rho _{\omega \ell }(r)Y_{\ell m}\left(
\theta ,\phi \right) e^{-i\omega t},  \label{curvedsolution}
\end{equation}%
where $\omega >0$, $Y_{\ell m}$ is the spherical harmonic function and the
radial function $\rho _{\omega \ell }$ satisfies
\begin{equation}
\frac{d^{2}\rho _{\omega \ell }}{dr^{\ast 2}}+\left\{ \omega ^{2}-\left( 1-%
\frac{2M}{r}\right) \left[ \frac{\ell (\ell +1)}{r^{2}}+\frac{2M}{r^{3}}%
\right] \right\} \rho _{\omega \ell }=0,  \label{re}
\end{equation}%
with $r^{\ast }$ being the tortoise coordinate defined as $r^{\ast
}=r+2M\log (r/2M-1)$. When $r\rightarrow \infty $, the radial function can
be solved as
\begin{equation}
\rho _{\omega \ell }^{r\rightarrow \infty }=e^{\pm i\omega r^{\ast }}.
\label{rf}
\end{equation}%
However, analytical solutions are not available at any finite radial
position, necessitating the use of numerical solutions. From the asymptotic
forms in Eq. (\ref{rf}), two types of field modes can be classified. One
type is the \textquotedblleft up-modes\textquotedblright\ associated with
mode solutions exhibiting the leading-order form $e^{+i\omega r^{\ast }}$ at
infinity. The other type is the \textquotedblleft
in-modes\textquotedblright\ associated with mode solutions exhibiting the
leading-order form $e^{-i\omega r^{\ast }}$ at infinity.

As outlined in Ref. \cite{Hodgkinson2014Static} (also refer to Appendix \ref{appendix A} for comprehensive calculation details), it is crucial to
handle the radial function with care. This is because the radial functions
solved from Eq. (\ref{re}) cannot be directly substituted into the mode
solution in Eq. (\ref{curvedsolution}) due to the differing boundary
conditions of the two modes.

In the exterior region of the Schwarzschild black hole, a complete set of
normalized basis functions for the massless scalar field is given by \cite%
{DeWitt1975ys,Candelas1980Vacuum}
\begin{eqnarray}
u_{\omega \ell m}^{in}(x) &=&\frac{1}{\sqrt{4\pi \omega }}\Phi _{\omega \ell
}^{in}(r)Y_{\ell m}(\theta ,\phi )e^{-i\omega t},  \notag \\
u_{\omega \ell m}^{up}(x) &=&\frac{1}{\sqrt{4\pi \omega }}\Phi _{\omega \ell
}^{up}(r)Y_{\ell m}(\theta ,\phi )e^{-i\omega t}.
\end{eqnarray}%
In this context, the radial function $\psi _{\omega \ell }(r)=\rho _{\omega
\ell }(r)/r$ of the mode solution in Eq. (\ref{curvedsolution}) is
normalized to yield the normalized function $\Phi _{\omega \ell }(r)$.
Utilizing the bases $u_{\omega \ell m}^{in}(x)$ and $u_{\omega \ell
m}^{up}(x)$, the quantum field can be expressed for the vacuum states within
the black hole spacetime. Further details can be found in the Appendix \ref{appendix B}.

The further calculation necessitates the determination of the detector's
geodesics, which remains the same for the two distinct vacuum states. We
consider the detector initiating its free-fall at a distance $R$ from the
black hole, with the geodesics given as \cite{chandrasekhar1985mathematical}
\begin{equation}
\left( \frac{dr}{d\tau }\right) ^{2}=\frac{2M}{r}-(1-E^{2}),\quad \frac{dt}{%
d\tau }=\frac{E}{1-2M/r}.  \label{wordline}
\end{equation}%
where $E^{2}=1-2M/R$ is related to the initial position $R$. In this paper,
we consider a finite free-falling distance $\Delta R$ for the detector.

Firstly, we calculate the transition amplitude for the Hartle-Hawking vacuum
state. The Hartle-Hawking vacuum state $|0\rangle _{H}$ is defined over the
entire spacetime and represents a state in which the black hole and its
surroundings are in thermal equilibrium at the same temperature as that of
the black hole \cite{Hartle1976Path,israel1976thermo}.

\begin{figure}[tbp]
\centering
\includegraphics[width=0.8\linewidth]{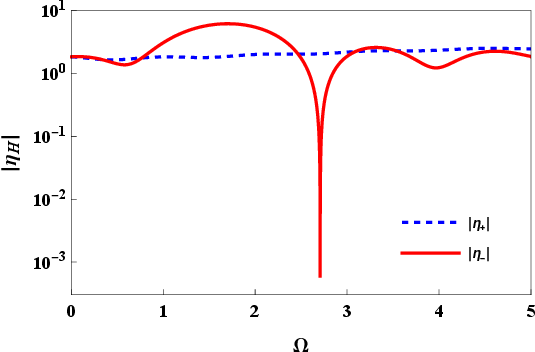}
\caption{The transition amplitude as a function of the atomic energy gap for
the Hartle-Hawking state. The red solid line denotes the process in which
the detector absorbs a photon from the field and then jumps from the ground
state to the excited state, and the blue dashed line denotes the process in
which the field excites a photon while the detector jumps to the excited
state.}
\label{Fig1}
\end{figure}

In order to calculate the transition amplitude in the Hartle-Hawking state,
we must expand the quantum field in terms
\begin{eqnarray}
w_{\omega \ell m}^{in} &=&\frac{1}{\sqrt{2\sinh \left( 4\pi M\omega \right) }%
}\left( e^{2\pi M\omega }u_{\omega \ell m}^{in}+e^{-2\pi M\omega }v_{\omega
\ell m}^{in\ast }\right) ,  \notag \\
\bar{w}_{\omega \ell m}^{in} &=&\frac{1}{\sqrt{2\sinh \left( 4\pi M\omega
\right) }}\left( e^{-2\pi M\omega }u_{\omega \ell m}^{in\ast }+e^{2\pi
M\omega }v_{\omega \ell m}^{in}\right) ,  \notag \\
w_{\omega \ell m}^{up} &=&\frac{1}{\sqrt{2\sinh \left( 4\pi M\omega \right) }%
}\left( e^{2\pi M\omega }u_{\omega \ell m}^{up}+e^{-2\pi M\omega }v_{\omega
\ell m}^{up\ast }\right) ,  \notag \\
\bar{w}_{\omega \ell m}^{up} &=&\frac{1}{\sqrt{2\sinh {(4\pi M\omega )}}}%
\left( e^{-2\pi M\omega }u_{\omega \ell m}^{up\ast }+e^{2\pi M\omega
}v_{\omega \ell m}^{up}\right) ,  \label{w}
\end{eqnarray}%
In the exterior region of the black hole, the $v$ functions vanish. With
these terms, we can expand the field as
\begin{eqnarray}
\phi _{H} &=&\sum_{\ell =0}^{\infty }\sum_{m=-\ell }^{+\ell
}\int_{0}^{\infty }d\omega (d_{\omega \ell m}^{up}w_{\omega \ell m}^{up}+%
\bar{d}_{\omega \ell m}^{up}\bar{w}_{\omega \ell m}^{up}  \notag \\
&&+d_{\omega \ell m}^{in}w_{\omega \ell m}^{in}+\bar{d}_{\omega \ell m}^{in}%
\bar{w}_{\omega \ell m}^{in}+H.c.)  \label{phiH}
\end{eqnarray}%
where $d^{j}$ and $\bar{d}^{j}$ are the annihilation operators, and satisfy
\begin{equation}
d_{\omega \ell m}^{j}|0\rangle _{H}=\bar{d}_{\omega \ell m}^{j}|0\rangle
_{H}=0.  \label{Hstate}
\end{equation}%
where $j\in \{in,up\}$. $H.c.$ represents the Hermitian conjugate terms and $%
d^{j\dagger }$ and $\bar{d}^{j\dagger }$ are the corresponding creation
operators.

Upon substituting the field $\phi _{H}$ into the Eqs. (\ref{etazf}) (see the Appendix \ref{appendix C} for the detailed calculation), we
derive the transition amplitude of the detector, as depicted in Fig. 1. The
figure vividly illustrates the variation of the transition amplitude.
Notably, a distinct absorption phenomenon emerges for a specific energy gap,
resembling the behavior observed in AIT. This observation indicates the
presence of gravity-induced transparency within the Schwarzschild spacetime
for the Hartle-Hawking state.

In contrast to the Hartle-Hawking state, which arises from the idealization
of time symmetry, the Unruh vacuum state is associated with a realistic
black hole formed via collapse and emerges from the inquiry into the reality
of black hole radiation. Consequently, the Unruh state is not applicable to
the in-modes originating from past infinity. Then, we can expand the field
\begin{eqnarray}
\phi _{U} &=&\sum_{\ell =0}^{\infty }\sum_{m=-\ell }^{+\ell
}\int_{0}^{\infty }d\omega (d_{\omega \ell m}^{up}w_{\omega \ell m}^{up}+%
\bar{d}_{\omega \ell m}^{up}\bar{w}_{\omega \ell m}^{up}  \notag \\
&&+b_{\omega \ell m}^{in}u_{\omega \ell m}^{in}+H.c.),
\end{eqnarray}%
where $b_{\omega \ell m}^{in}|0\rangle _{U}=d_{\omega \ell m}^{up}|0\rangle
_{U}=\bar{d}_{\omega \ell m}^{up}|0\rangle _{U}=0$.

\begin{figure}[tbp]
\centering
\includegraphics[width=0.8\linewidth]{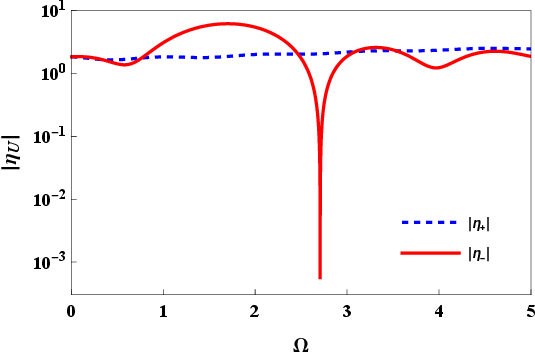}
\caption{Transition amplitude of the detector for the Unruh state. The lines
in this figure have the same meaning as that in Fig. 1.}
\label{Fig2}
\end{figure}

Similar to the calculation for the Hartle-Hawking state, we can obtain the
results of the gravity-induced transparency phenomenon in the Unruh state,
as illustrated in Fig. 2. It is found that the energy gap required for the
emergence of gravity-induced transparency is essentially the same as that in
the Hartle-Hawking state. This similarity arises from the fact that the
field types used in defining the Unruh and Hartle-Hawking states are
fundamentally alike. For both the Unruh state and the Hartle-Hawking state,
the transition rates of the detector are proportional to $\frac{1}{%
2E(e^{E/kT}-1)}$ and $\frac{1}{E(e^{E/kT}-1)}$, respectively \cite%
{birrell1984quantum}, differing only by a constant factor of $1/2$. Hence,
the change in the transition amplitude with the energy gap of the detector
also yields essentially the same results.

Finally, it's worth emphasizing that the Boulware vacuum state \cite{dgb1975}
is not suitable for our study. This is because no Hawking radiation exists
relative to the Boulware vacuum for the free-falling observers. In
particular, the Boulware vacuum is not employed for the free-falling
observers due to the divergence of the energy-momentum tensor at the horizon
of the black hole \cite{birrell1984quantum}.

\section{Conclusion}

In this paper, we employ numerical methods to investigate a detector coupled
to a massless scalar field in both the Hartle-Hawking and Unruh states. We
compute the transition amplitudes of the detector in these two states and
identify gravity-induced transparency phenomena, akin to AIT phenomena
observed for an accelerated detector in an electromagnetic field. Notably,
we find that the energy gap of the detector remains consistent between the
two states when gravity-induced transparency occurs. This consistency arises
from the nearly identical definitions of field modes in these two states for
our purpose. Extending our study to other black hole spacetimes presents an
intriguing avenue for future research.

\section{Acknowledgments}

This work is supported by National Natural Science Foundation of China
(NSFC) with Grant No. 12375057 and the Fundamental Research Funds for the
Central Universities, China University of Geosciences (Wuhan).

\appendix

\section{\textbf{Radial Function} $\protect\rho _{\protect\omega \ell }$}\label{appendix A}

To solve the differential equation in Eq. (11) of the main text, we
introduce the function
\begin{equation}
\psi _{\omega \ell }(r):=\rho _{\omega \ell }(r)/r,  \label{rhoa1}
\end{equation}%
which satisfies:
\begin{equation}\label{psi}
	\begin{aligned}
		\psi _{\omega \ell }^{\prime \prime }(r)&+\frac{2(r-M)}{r(r-2M)}\psi _{\omega
			\ell }^{\prime }(r)
		\\&+\left( \frac{\omega ^{2}r^{2}}{(r-2M)^{2}}-\frac{\xi }{%
			r(r-2M)}\right) \psi _{\omega \ell }(r)=0,
	\end{aligned}
\end{equation}%
with $\xi =\ell (\ell +1).$

For the up-modes, to numerically obtain their value at a given suitably
large radius $r_{\infty }$, we substitute the ansatz
\begin{equation}
\psi _{a\ell }^{up}\sim \frac{e^{iar^{\ast }}}{r}e^{v(r)},\qquad
v(r):=\sum_{n=1}^{\infty }\frac{c_{n}}{r^{n}},  \label{vr}
\end{equation}%
into Eq. (\ref{psi}) and leads to an equation for $v(r)$,

\begin{equation}
	\begin{aligned}
		&r^{2}(r-2M)v^{\prime \prime }(r)+r^{2}(r-2M)(v^{\prime
		}(r))^{2}
	  \\&+2r(M+i\omega r^{2})v^{\prime }(r)-(\ell (\ell +1)r+2M)=0.
	\end{aligned}
\label{vreq}
\end{equation}%
Then, we substitute the expression of $v(r)$ in Eq. (\ref{vr}) into Eq. (\ref%
{vreq}) and collect inverse powers of $r$. The coefficient of each power of $%
r$ must be set equal to zero. In practice, the upper limit in the sum (\ref%
{vr}) is replaced by some suitable cutoff, denoted as $n_{\infty }$.

The initial conditions for the up-modes are taken as
\begin{equation}
	\begin{aligned}
		&\psi _{\omega \ell }^{up}(r_{\infty })=\frac{e^{i\omega r^{\ast }(r_{\infty })}}{r_{\infty }}e^{v(r_{\infty })},\text{ }
		\\&\psi _{a\ell }^{up\prime
		}(r_{\infty })=\frac{d}{dr}\left[ \frac{e^{i\omega r^{\ast }(r)}}{r}e^{v(r)}%
		\right] _{r=r_{\infty }}.  \label{ic}
	\end{aligned}
\end{equation}%
These conditions would become more accurate as $n_{\infty }$ and $r_{\infty
} $ take larger values. We use the initial conditions (\ref{ic}) to make the
calculation and take the parameters as $n_{\infty }=100$ and $r_{\infty
}=15000M$, and thus we can obtain a numerical solution for $\psi _{\omega
\ell }^{up}(r)$.

For the in-modes, we have
\begin{equation}
\psi _{\omega \ell }^{in}\sim \frac{e^{-i\omega r^{\ast }}}{r}w(r),\qquad
w(r):=\sum_{n=0}^{\infty }b_{n}(r-2M)^{n}.  \label{wr}
\end{equation}%
Similarly, we substitute Eq. (\ref{wr}) into Eq. (\ref{psi}) to obtain an
equation about $w(r)$ as
\begin{equation}\label{wre}
	\begin{aligned}
		r^{2}(r-2M)w^{\prime \prime }(r)&+2r(M-ir^{2}\omega )w^{\prime }(r)\\
		&-(\ell
		(\ell +1)r+2M)w(r)=0.
	\end{aligned}
\end{equation}%
According to Eq. (\ref{wr}) and (\ref{wre}), we can get%
\begin{eqnarray*}
b_{0}&=&1,\quad b_{-1}=b_{-2}=0, \\
b_{n}&=&\frac{-[-12i\omega M(n-1)+(2n-3)(n-1)-(\ell (\ell +1)+1)]}{%
2M(n^{2}-i4Mn\omega )}b_{n-1} \\
&&-\frac{[(n-2)(n-3)-i12M\omega (n-2)-\ell (\ell +1)]}{4M^{2}(n^{2}-i4Mn%
\omega )}b_{n-2}\\
&&+\frac{i\omega (n-3)}{2M^{2}(n^{2}-i4Mn\omega )}b_{n-3}.
\end{eqnarray*}

The initial conditions for the in-modes are taken as
\begin{equation}\label{ici}
	\begin{aligned}
		&\psi _{\omega \ell }^{in}(r_{H})=\frac{e^{-i\omega r^{\ast }}(r_{H})}{r_{H}}%
		w(r_{H}),\\
		&\psi _{\omega \ell }^{in\prime }(r_{H})=\frac{d}{dr}\left[
		\frac{e^{-i\omega r^{\ast }(r)}}{r}w(r)\right] _{r=r_{H}}.
	\end{aligned}
\end{equation}%
We use the initial conditions (\ref{ici}) to make the calculation and take
the parameters as $n_{\infty }=200$ and $r_{H}=2.0000001M$, and thus we can
obtain a numerical solution for $\psi _{\omega \ell }^{in}(r)$.

The numerical results for $\psi _{\omega \ell }^{up}(r)$ and $\psi _{\omega
\ell }^{in}(r)$ can be used to calculate the transition amplitude of the
detector in the curved spacetime.

\section{Normalisation}\label{appendix B}

In this section we deal with the normalization of radial functions for the
mode solution in Eq. (10) of the main text.

We choose radial function $\Psi $ satisfies the following asymptotic form
\begin{eqnarray}
\Psi _{\omega \ell }^{in}(r) &\sim &\{%
\begin{matrix}
B_{\omega \ell }^{in}e^{-i\omega r^{\ast }}, & r\rightarrow 2M, \\
r^{-1}e^{-i\omega r^{\ast }}+A_{\omega \ell ^{\prime
}}^{in}r^{-1}e^{+i\omega r^{\ast }}, & r\rightarrow \infty ,%
\end{matrix}
\notag \\
&&  \notag \\
\Psi _{\omega \ell }^{up}(r) &\sim &\{%
\begin{array}{ll}
A_{\omega \ell }^{up}e^{-i\omega r^{\ast }}+e^{+i\omega r^{\ast }}, &
r\rightarrow 2M, \\
B_{\omega \ell }^{up}r^{-1}e^{+i\omega r^{\ast }}, & r\rightarrow \infty .%
\end{array}
\label{asymptotic}
\end{eqnarray}%
and $A_{\omega \ell }^{up},B_{\omega \ell }^{up}$ are the transmission and
reflection coefficients that satisfy the following Wronskian relations
\begin{eqnarray}
&&B_{\omega \ell }^{up}=\frac{(2M)2i\omega }{W[\rho _{\omega \ell
}^{in},\rho _{\omega \ell }^{up}]},\quad A_{\omega \ell }^{up}=-\frac{W[\rho
_{\omega \ell }^{in},\rho _{\omega \ell }^{up}]^{\ast }}{W[\rho _{\omega
\ell }^{in},\rho _{\omega \ell }^{up}]},  \notag \\
&&A_{\omega \ell }^{in}=-\frac{W[\rho _{\omega \ell }^{in},\rho _{\omega
\ell }^{up\ast }]}{W[\rho _{\omega \ell }^{in},\rho _{\omega \ell }^{up}]}.
\label{coefficients}
\end{eqnarray}%
We replace the function $\Psi _{\omega \ell }^{j}(r)$ with $F_{\omega \ell
}^{j}(r)$, $j\in \{in,up\}$, defined by
\begin{equation}
F_{\omega \ell }^{in}=\Psi _{\omega \ell }^{in},\quad F_{\omega \ell }^{up}=%
\frac{\Psi _{\omega \ell }^{up}}{2M}.  \label{leibi}
\end{equation}%
It can be verified that the functions $R_{\omega \ell }^{j}(r):=rF_{\omega
\ell }^{j}(r)$ are normalized, with the relationship between reflection and
transmission coefficients
\begin{eqnarray}
B_{a\ell \ell }^{up}=(2M)^{2}B_{a\ell \ell }^{in},\quad &|A_{\omega \ell
}^{in}|^{2}&=1-4M^{2}|B_{\omega \ell }^{in}|^{2},  \notag \\
|A_{a\ell }^{in}|^{2}=|A_{a\ell }^{up}|^{2},\quad &|A_{a\ell }^{up}|^{2}&=1-%
\frac{|B_{\omega \ell }^{up}|^{2}}{4M^{2}}.  \label{rs}
\end{eqnarray}

Then, we can define
\begin{eqnarray}
\Phi _{\omega \ell }^{in}(r) &=&\dfrac{B_{\omega \ell }^{up}\psi _{\omega
\ell }^{in}(r)}{2M},  \notag \\
\Phi _{\omega \ell }^{up}(r) &=&\dfrac{B_{\omega \ell }^{up}\psi _{\omega
\ell }^{up}(r)}{2M}.  \label{PSI}
\end{eqnarray}%
Using the Eq. (\ref{rs}), it can be shown that the solution in Eq. (13) of
the main text satisfies the orthogonal normalization relation as follows%
\begin{eqnarray}
(u_{\omega tm}^{up},u_{\omega ^{\prime }t^{\prime }m^{\prime }}^{up})
&=&\delta _{\ell \ell ^{\prime }}\delta _{mm^{\prime }}\delta (\omega
-\omega ^{\prime }),  \notag \\
(u_{\omega \ell m}^{in},u_{\omega ^{\prime }\ell ^{\prime }m^{\prime
}}^{in}) &=&\delta _{\ell \ell ^{\prime }}\delta _{mm^{\prime }}\delta
(\omega -\omega ^{\prime }),  \notag \\
(u_{\omega \ell m}^{in},u_{\omega ^{\prime }\ell ^{\prime }m^{\prime
}}^{up}) &=&0,
\end{eqnarray}%
and the coefficient $B_{\omega \ell }^{up}$ can be obtained by substituting $%
\rho _{\omega \ell }^{up}$ and $\rho _{\omega \ell }^{in}$ into (\ref%
{coefficients}).

\section{Transition Amplitude}\label{appendix C}

We consider free-falling process of the detector starting at the position $R$
which is the distance from the center of the black hole.

For the Hartler-Hawking state, the transition amplitude is calculated as%
\begin{eqnarray}
\eta _{H}^{+} &=&\int d\tau e^{i\Omega \tau }\langle n+1|_{H}\phi _{H}(\tau
)|n\rangle _{H}  \notag \\
&=&\int d\tau e^{i\Omega \tau }\langle n+1|_{H}\sum_{\ell =0}^{\infty
}\sum_{m=-\ell }^{+\ell }\int_{0}^{\infty }d\omega (d_{\omega \ell
m}^{up}w_{\omega \ell m}^{up}  \notag \\
&&+\bar{d}_{\omega \ell m}^{up}\bar{w}_{\omega \ell m}^{up}+d_{\omega \ell
m}^{in}w_{\omega \ell m}^{in}+\bar{d}_{\omega \ell m}^{in}\bar{w}_{\omega
\ell m}^{in}+H.c.)|n\rangle _{H}  \notag \\
&=&\int d\tau e^{i\Omega \tau }\sum_{\ell =0}^{\infty }\sum_{m=-\ell
}^{+\ell }\int_{0}^{\infty }d\omega \notag\\
&& (w_{\omega \ell m}^{up\ast }+\bar{w}%
_{\omega \ell m}^{up\ast }+w_{\omega \ell m}^{in\ast }+\bar{w}_{\omega \ell
m}^{in\ast })  \notag \\
&=&f_{1}+f_{2}+f_{3}+f_{4}.
\end{eqnarray}%
Note that the integration over field frequencies will vanish when we only
consider a single field frequency. The above equation is divided into four
terms and their calculation process is similar. Here, we take the
calculation process for $f_{1}$ as an example,
\begin{eqnarray}
f_{1} &=&\int d\tau e^{i\Omega \tau }\sum_{\ell =0}^{\infty }\sum_{m=-\ell
}^{+\ell }(w_{\omega \ell m}^{up\ast })  \notag \\
&=&\int d\tau e^{i\Omega \tau }\sum_{\ell =0}^{\infty }\sum_{m=-\ell
}^{+\ell }\frac{1}{\sqrt{2\sinh \left( 4\pi \omega M\right) }}(e^{2\pi
M\omega }u_{\omega \ell m}^{up\ast })  \notag \\
&=&\int d\tau e^{i\Omega \tau }\sum_{\ell =0}^{\infty }\sum_{m=-\ell
}^{+\ell }\frac{e^{2\pi \omega M}Y_{\ell m}^{\ast }(\theta ,\beta
)e^{i\omega t}}{\sqrt{8\pi \omega \sinh \left( 4\pi M\omega \right) }}\Phi
_{\omega \ell }^{up\ast }(r)  \notag \\
&=&\int d\tau e^{i\Omega \tau }\sum_{\ell =0}^{\infty }\sum_{m=-\ell
}^{+\ell }\frac{e^{2\pi \omega M}Y_{\ell m}^{\ast }(\theta ,\beta
)e^{i\omega t}}{\sqrt{8\pi \omega \sinh \left( 4\pi M\omega \right) }}\dfrac{%
B_{\omega \ell }^{up}\psi _{\omega \ell }^{up\ast }(r)}{2M}
\end{eqnarray}

For a detector that starts its free fall at $R$, we cannot obtain a precise
expression for the world line $r(\tau ),t(\tau )$ from Eq. (14) of the main
text. So we can't just substitute this world line into the field $\phi
_{S}[t(\tau ),r(\tau )]$. We require to make the transformation to obtain
the geodesics,
\begin{eqnarray}
d\tau &=&\frac{dr}{\sqrt{\frac{2M}{r}-(1-E^{2})}},  \notag \\
dt &=&\frac{E}{1-2M/r}\frac{dr}{\sqrt{\frac{2M}{r}-(1-E^{2})}}.  \label{wi}
\end{eqnarray}%
In particular, $E=1$ when the detector starts its free fall from infinity $%
R=\infty $. For our purpose, the detector starts its free fall from $%
2M<R=Constant$. Make the integral in Eq. (\ref{wi}) to obtain,
\begin{eqnarray}
\tau (r) &=&\frac{\sqrt{-1+E^{2}+\frac{2M}{r}}}{-1+E^{2}}-\frac{2M\text{ tan}%
^{-1}\left[ \frac{\sqrt{-1+E^{2}+\frac{2M}{r}}}{\sqrt{1-E^{2}}}\right] }{%
(1-E^{2})^{3/2}},  \notag \\
t(r) &=&\frac{E\sqrt{-1+E^{2}+\frac{2N}{r}}}{-1+E^{2}}-4M\mathrm{\tanh }^{-1}\left[ \frac{\sqrt{-1+E^{2}+\frac{2N}{r}}}{E}\right]  \notag \\
&&+E\frac{2\left(
	-3+2E^{2}\right) M\text{tan}^{-1}\left[ \frac{\sqrt{-1+E^{2}+\frac{2N}{r}}}{%
		\sqrt{1-E^{2}}}\right] }{(1-E^{2})^{3/2}}
\end{eqnarray}

Then, the function $f_{1}$ becomes
\begin{equation}
	\begin{aligned}
		f_{1}=&\int_{R-\Delta R}^{R}dr\sum_{\ell =0}^{\infty }\sum_{m=-\ell }^{+\ell }%
		\frac{e^{i\Omega \tau (r)}}{\sqrt{\frac{2M}{r}-\frac{2M}{R}}}\\
		&\times\frac{e^{2\pi
				\omega M}Y_{\ell m}^{\ast }(\theta ,\beta )e^{i\omega t(r)}}{\sqrt{8\pi
				\omega \sinh \left( 4\pi M\omega \right) }}\dfrac{B_{\omega \ell }^{up}\psi
			_{\omega \ell }^{up\ast }(r)}{2M},
	\end{aligned}
\end{equation}%
where $\Delta R$ is the free-falling distance for the detector. To get
around the divergence caused by the term $1/(\sqrt{2M/r-2M/R})$, we change
the integral over $R$ into the form of a discrete point summation and remove
the initial point. Thus, we get%
\begin{equation}
	\begin{aligned}
		f_{1}=&\sum_{n=0}^{n=N}\sum_{\ell =0}^{\infty }\sum_{m=-\ell }^{+\ell }\frac{%
			e^{i\Omega \tau (r_{n})}}{\sqrt{\frac{2M}{r_{n}}-\frac{2M}{R}}}\\
		&\times\frac{e^{2\pi
				\omega M}Y_{\ell m}^{\ast }(\theta ,\beta )e^{i\omega t(r_{n})}}{\sqrt{8\pi
				\omega \sinh \left( 4\pi M\omega \right) }}\dfrac{B_{\omega \ell }^{up}\psi
			_{\omega \ell }^{up\ast }(r_{n})}{2M},
	\end{aligned}
\end{equation}%
where $r_{n}$ denotes the $n$-th value in the $r_{n}\in \{R-\Delta
R,R-\varepsilon \}$ interval up to $N$ points. In the concrete calculation,
we take the cutoff of $\ell $ to be $\ell =5$, and the parameters $N=20$, $%
R=4.0029299M$, $\Delta R=1.0029299M$, $\varepsilon =0.0029299$.

The remaining three terms, $f_{2},f_{3},f_{4}$, have a similar calculation.
And the transition amplification in Eq. (9) of the main text for the Unruh
vacuum state is calculated in the same way as $\eta _{H}^{+}$.

\end{document}